\documentclass[final,5p,times,twocolumn]{elsarticle}

\usepackage{amssymb}
\usepackage{graphicx}
\usepackage{pzccal}
\usepackage{color}

\newcommand{\eqref}[1]{(\ref{#1})}

\biboptions{sort&compress}

\journal{Physics Letters B}

\begin{document}
\begin{frontmatter}

\title{Nucleon spin structure at very high-$x$}

\author[ANL]{Craig D.~Roberts}
\author[ANL]{Roy J.~Holt}
\author[JARA]{Sebastian M.~Schmidt}

\address[ANL]{Physics Division, Argonne National Laboratory, Argonne, Illinois 60439, USA}
\address[JARA]{Institute for Advanced Simulation, Forschungszentrum J\"ulich and JARA, D-52425 J\"ulich, Germany}

\date{\today}


\begin{abstract}
Dyson-Schwinger equation treatments of the strong interaction show that the presence and importance of nonpointlike diquark correlations within the nucleon are a natural consequence of dynamical chiral symmetry breaking.  Using this foundation, we deduce a collection of simple formulae, expressed in terms of diquark appearance and mixing probabilities, from which one may compute ratios of longitudinal-spin-dependent $u$- and $d$-quark parton distribution functions on the domain $x\simeq 1$.  A comparison with predictions from other approaches plus a consideration of extant and planned experiments shows that the measurement of nucleon longitudinal spin asymmetries on $x\simeq 1$ can add considerably to our capacity for discriminating between contemporary pictures of nucleon structure.
\end{abstract}


\end{frontmatter}

%
\hspace*{-\parindent}\textbf{1 Introduction}.  Since the advent of the parton model and the first deep inelastic scattering (DIS) experiments there has been a determined effort to deduce the parton distribution functions (PDFs) of the most stable hadrons: neutron, proton and pion \cite{Holt:2010vj}.  The behavior of such distributions on the far valence domain (Bjorken-$x> 0.5$) is of particular interest because this domain is definitive of hadrons; e.g., quark content on the far valence domain is how one distinguishes between a neutron and a proton.  Indeed, all Poincar\'e-invariant properties of a hadron: baryon number, charge, flavour content, total spin, etc., are determined by the PDFs which dominate on the far valence domain.

Recognizing the significance of the far valence domain, a new generation of experiments, focused on $x\gtrsim 0.5$, is planned at the Thomas Jefferson National Accelerator Facility (JLab) \cite{Dudek:2012vr}, and under examination in connection with Drell-Yan studies at the Fermi National Accelerator Facility (FNAL) \cite{Reimer:Figure} and a possible Electron Ion Collider (EIC), in China or the USA.  Consideration is also being given to experiments aimed at measuring parton distribution functions in mesons at the Japanese Proton Accelerator Research Center facility (J-PARC).  Furthermore, at the Facility for Antiproton and Ion Research (FAIR), under construction in Germany, it would be possible to directly measure the Drell-Yan process from high-$x$ antiquarks in the antiproton annihilating with quarks in the proton.  A spin physics program at the Nuclotron based Ion Collider fAcility (NICA), under development in Dubna, might also make valuable contributions.

A concentration on such measurements demands that theory move beyond merely parametrising PDFs (and parton distribution amplitudes, too \cite{Chang:2013pq,Cloet:2013tta,Chang:2013nia}).  Computation within frameworks with a traceable connection to QCD becomes critical because without it, no amount of data will reveal anything about the theory underlying strong interaction phenomena.  This is made clear by the example of the pion's valence-quark PDF, $u_v^\pi(x)$, in connection with which a failure of QCD was suggested following a leading-order analysis of $\pi N$ Drell-Yan measurements \cite{Conway:1989fs}.  As explained in Ref.\,\cite{Holt:2010vj}, this confusion was fostered by the application of a diverse range of models.  On the other hand, a series of QCD-connected calculations  \cite{Hecht:2000xa,Wijesooriya:2005ir,Aicher:2010cb,Nguyen:2011jy} subsequently established that the leading-order analysis was misleading, so that $u_v^\pi(x)$ may now be seen as a success for the unification of nonperturbative and perturbative studies in QCD.

The endpoint of the far valence domain, $x=1$, is especially significant because, whilst all familiar PDFs vanish at $x=1$, ratios of any two need not; and, under DGLAP evolution, the value of such a ratio is invariant \cite{Holt:2010vj}.  Thus, e.g., with $d_v(x)$, $u_v(x)$ the proton's $d$, $u$ valence-quark PDFs, the value of $\lim_{x\to 1} d_v(x)/u_v(x)$ is an unambiguous, scale invariant, nonperturbative feature of QCD.  It is therefore a keen discriminator between frameworks that claim to explain nucleon structure.  Furthermore, Bjorken-$x=1$ corresponds strictly to the situation in which the invariant mass of the hadronic final state is precisely that of the target; viz., elastic scattering.  The structure functions inferred experimentally on the neighborhood $x\simeq 1$ are therefore determined theoretically by the target's elastic form factors.

One may contrast these favorable circumstances with the situation encountered when attempting to distinguish between PDF computations on $x\lesssim 0.85$.  Diverse models produce PDFs with markedly different profiles on this domain.  However, practitioners then augment their computation with a statement that the result is valid at a ``model scale''; i.e., an \emph{a priori} unknown momentum scale, $\zeta_0$, which is treated as a parameter.  This parameter is subsequently chosen to be that momentum-scale required as the starting point for DGLAP-evolution in order to obtain agreement, according to some subjective criteria, with a modern PDF parametrization at some significantly larger scale: $\zeta/\zeta_0\gtrsim 10$.  This procedure serves to diminish any capacity for discriminating between models.  Herein, we therefore focus on predictions for PDF ratios on $x\simeq 1$.
\smallskip

%
\hspace*{-\parindent}\textbf{2 Faddeev equation}.  The connection between $x=1$ and hadron elastic form factors provides a direct link between computations of a nucleon's Poincar\'e covariant Faddeev amplitude and predictions for the $x=1$ value of PDF ratios \cite{Holt:2010vj}.  The amplitude is obtained from a Faddeev equation, which is one of the collection of Dyson-Schwinger equations (DSEs) \cite{Bashir:2012fs}.  In composing the Faddeev equation, one begins with dressed-quark propagator, which is obtained from QCD's gap equation:
\begin{eqnarray}
\nonumber
\lefteqn{
S(p)^{-1} = i\gamma\cdot p + m}\\
&&
+ \int\frac{d^4q}{(2\pi)^4} \, g^2 D_{\mu\nu}(p-q)\frac{\lambda^a}{2}\gamma_\mu S(q) \frac{\lambda^a}{2}\Gamma_\nu(q,p) , \rule{1em}{0ex}
\label{gendseN}
\end{eqnarray}
wherein $D_{\mu\nu}(k)=[\delta_{\mu\nu}-k_\mu k_\nu/k^2]\Delta(k^2)$ is the gluon propagator; $\Gamma_\nu$, the quark-gluon vertex; and $m$, the current-quark bare mass.  (Renormalisation is discussed elsewhere \cite{Maris:1997tm}.)  The dynamical content of the kernel in Eq.\,\eqref{gendseN} is understood.  The gluon propagator may be obtained from its own gap equation; and modern studies  \cite{Boucaud:2011ug,Ayala:2012pb} show that $\Delta(k^2)$ is a bounded, regular, monotonic function of spacelike momenta, which achieves its maximum value on this domain at $k^2=0$.  Moreover, the dressed-quark--gluon vertex does not possess any structure that can qualitatively alter this behavior \cite{Skullerud:2003qu,Bhagwat:2004kj,He:2013jaa}.

The gap equation's solution is the dressed-quark propagator:
\begin{equation}
S(p) 
=Z(p^2)/[i\gamma\cdot p  + M(p^2)]\,,
\label{eqSp}
\end{equation}
where $Z(p^2)$ is the wave-function renormalisation and $M(p^2)$ is the dressed-quark mass-function.  In QCD with massless current-quarks, any finite-order perturbative computation yields $M(p^2)\equiv 0$.  However, a nonperturbative solution of Eq.\,\eqref{gendseN} predicts a nonzero mass function with a strong momentum dependence \cite{Roberts:2007ji}.  This prediction is confirmed by simulations of lattice-QCD \cite{Bowman:2005vx}, so that it is now theoretically established that chiral symmetry is dynamically broken in QCD \cite{national2012Nuclear}.  It follows that dressed light-quarks are characterised by an infrared ``spectrum mass'' of $M(p^2=0)\approx 0.4\,$GeV$=:M_D\,$, with the behaviour of perturbative QCD recovered for $p^2\gtrsim 4\,$GeV$^2$.

\begin{figure}[t]
\centerline{%
\includegraphics[clip,width=0.47\textwidth]{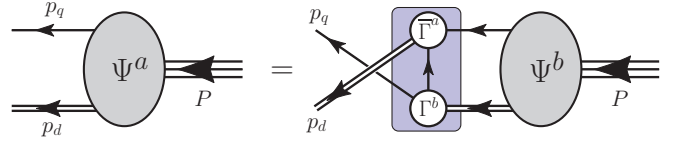}}
\caption{\label{figFaddeev} Poincar\'e covariant Faddeev equation.  $\Psi$ is the Faddeev amplitude for a proton of total momentum $P= p_q + p_d$.  The shaded rectangle demarcates the kernel of the Faddeev equation: \emph{single line}, dressed-quark propagator; $\Gamma$, diquark correlation amplitude; and \emph{double line}, diquark propagator.}
\end{figure}

With a solution to the one-body problem in hand, bound-states can be considered; and in a DSE treatment of the meson sector the focus is on symmetry-preserving analyses of the Bethe-Salpeter equation (BSE) for quark-antiquark vertices and bound-states \cite{Chang:2011vu}.  The baryonic analogue is the Faddeev equation, which may be derived by considering that a baryon appears as a pole in a six-point quark Green function, with the residue proportional to the baryon's Faddeev amplitude.  The Faddeev equation then sums all possible exchanges and interactions that can take place between three dressed-quarks.  This was first considered in Ref.\,\cite{Cahill:1988dx}, wherein a tractable simplification was presented.  Namely, as a dynamical consequence of strong binding in the colour singlet meson sector, exposed through BSE studies, it was shown \cite{Cahill:1987qr} that the quark$+$quark$\to$quark$+$quark scattering matrix, ${\cal M}_{qq}$, which appears as a component of the Faddeev equation, is accurately approximated by a sum of nonpointlike quark$+$quark (diquark) correlations in the colour-antitriplet channel (Eq.\,(A.27) in Ref.\,\cite{Cloet:2008re}).  This is an immense simplification because it reduces the three-body problem to an equation that is essentially two-body in nature; viz., Fig.\,\ref{figFaddeev}.

The preceding material highlights that diquark correlations are not inserted into the Faddeev equation ``by hand.''  Their appearance and importance are dynamical consequences of QCD's strong coupling and a further manifestation of the crucial role of DCSB \cite{Chen:2012qr}.  Whether one exploits this feature in approximating ${\cal M}_{qq}$ in the Faddeev equation \cite{Cloet:2007piS,Cloet:2008re,Eichmann:2008ef,Chang:2011tx,Wilson:2011aa,%
Cloet:2011qu,Chen:2012qr,Segovia:2013uga}
or eschews the simplification it offers, the outcome, when known, is the same \cite{Eichmann:2009qa}.  Empirical evidence supporting the presence of diquarks in the nucleon is accumulating \cite{Cates:2011pz,Wilson:2011aa,Cloet:2012cy,Qattan:2012zf}.  Furthermore, these dynamically generated correlations should not be confused with the pointlike diquarks introduced in order to simplify the study of systems constituted from three constituent-quarks \cite{Lichtenberg:1967zz,Lichtenberg:1968zz}.  The modern dynamical diquark correlation is nonpointlike, with the charge radius of a given diquark being typically 10\% larger than its mesonic analogue \cite{Roberts:2011wy}.  Hence, diquarks are soft components within baryons.

Detailed studies of the Faddeev equation in Fig.\,\ref{figFaddeev} have shown \cite{Cloet:2007piS,Cloet:2008re,Eichmann:2008ef,Chang:2011tx,Wilson:2011aa,Cloet:2011qu,Chen:2012qr,Segovia:2013uga}
that the dominant correlations for ground-state octet baryons are scalar ($0^+$) and axial-vector ($1^+$) diquarks because, e.g., the associated mass-scales are smaller than the baryons' masses and their parity matches that of these baryons.  Only $0^+$ and $1^+$ correlations need therefore be retained in approximating ${\cal M}_{qq}$.  A particular strength of the DSE approach is that it allows one to treat mesons and baryons on the same footing and, in particular, enables the impact of dynamical chiral symmetry breaking (DCSB), the origin of more than 98\% of the visible mass in the universe \cite{Bashir:2012fs,national2012Nuclear}, to be expressed in the prediction of baryon properties.  Notably, the most fundamental expression of DCSB in QCD is the momentum dependence of the dressed-quark mass-function \cite{Roberts:2007ji}; which has observable impacts \cite{Chang:2013pq,Cloet:2013tta,Chang:2013nia,Cloet:2013gva}.  On the other hand, for processes involving probe momentum-scales $Q\lesssim M_D$ the momentum-dependence is invisible and hence a vector$\,\otimes\,$vector contact interaction, which yields $M(p^2)=\,$constant$\,=M_D$ via Eq.\,\eqref{gendseN}, provides a reliable, predictive tool \cite{Roberts:2011wy,Wilson:2011aa,Chen:2012qr,Chen:2012txa,Segovia:2013uga}.  Herein, therefore, we compare DSE results obtained with both a realistic interaction, which produces a momentum-dependent dressed-quark mass, and a contact interaction, which gives $M(p^2)=M_D$.  This enables us to highlight the sensitivity of empirical observables to the infrared behaviour of QCD's running coupling.
\smallskip

\hspace*{-\parindent}\textbf{3 Theoretical Predictions}.  The connection between the \mbox{$Q^2=0$} values of elastic form factors and the behavior of PDFs in the neighborhood of $x=1$ has previously been exploited in the calculation of $d_v/u_v|_{x=1}$ \cite{Holt:2010vj,Wilson:2011aa}.  Since similar arguments will be used herein for polarised distributions, we will recapitulate upon that analysis.  The ratio $d_v/u_v|_{x=1}$ expresses the relative probability of finding a $d$-quark carrying all the proton's light-front momentum compared with that of a $u$-quark doing the same or, equally, owing to invariance under evolution, the relative probability that a $Q^2=0$ probe either scatters from a $d$- or a $u$-quark; viz.,
$d_v/u_v|_{x\simeq 1} = P_{1}^{p,d}/P_{1}^{p,u}$.
%
When a Poincar\'e-covariant Faddeev equation is employed to describe the nucleon, then
\begin{equation}
\label{dvuvF1result}
\left. \frac{d_v(x)}{u_v(x)}\right|_{x\simeq 1} = \frac{P_{1}^{p,d}}{P_{1}^{p,u}} =
\frac{\frac{2}{3} P_1^{p,a} + \frac{1}{3} P_1^{p,m}}
{P_1^{p,s}+\frac{1}{3} P_1^{p,a} + \frac{2}{3} P_1^{p,m}},
\end{equation}
where we have used the notation of Ref.\,\cite{Cloet:2008re}.  Namely,
$P_1^{p,s}=F_{1p}^s(Q^2=0)$ is the contribution to the proton's charge arising from diagrams with a scalar diquark in both the initial and final state: $u[ud]\otimes \gamma \otimes u[ud]$.  The diquark-photon interaction is far softer than the quark-photon interaction and hence this diagram contributes solely to $u_v$ at $x=1$.
$P_1^{p,a}$ is the kindred axial-vector diquark contribution: $2 d\{uu\}\otimes \gamma\otimes d\{uu\}+u\{ud\} \otimes\gamma \otimes u\{ud\}$.  At $x=1$ this contributes twice as much to $d_v$ as it does to $u_v$.
$P_1^{p,m}$ is the contribution to the proton's charge arising from diagrams with a different diquark component in the initial and final state.  The existence of a hard component in this contribution relies on the exchange of a quark between the diquark correlations and hence it contributes twice as much to $u_v$ as it does to $d_v$.  

It is plain from Eq.\,\eqref{dvuvF1result} that $d_v/u_v|_{x=1}=0$ in the absence of axial-vector diquark correlations; i.e., in scalar-diquark-only models of the nucleon \cite{Close:1988br}.  A context for this result is provided by the following observations: QCD predicts that at $\zeta\simeq 1\,$GeV the distribution of a valence-quark behaves as \cite{Ezawa:1974wm}
\begin{equation}
q_v(x) \stackrel{x\simeq 1}{\propto} (1-x)^{3+\gamma},
\end{equation}
where $0<\gamma\ll 1$ is an anomalous dimension, which grows under DGLAP evolution to larger scales; and the elastic electromagnetic form factor of a scalar diquark correlation behaves as $1/Q^2$ for large momentum transfers.  In this case, $d_v/u_v|_{x=1}=0$ entails that $d_v(x) \propto (1-x)^2 u_v(x)$ on $x\simeq 1$: the $d$-quark is sequestered within a soft diquark correlation and therefore plays no role in hard processes involving the proton.  Moreover, the pointwise behavior of $d_v(x)$ is plainly not that of a valence quark so there is a clear sense in which the result $d_v/u_v|_{x=1}=0$ means there are no truly-valence $d$-quarks in the proton.  In this case the degrees of freedom within the proton on the far valence domain are a $u$-quark and a soft scalar-diquark correlation.

It should be noted that any self-consistent solution of the Faddeev equation in Fig.\,\ref{figFaddeev} will produce a nucleon amplitude that contains axial-vector diquark components in addition to the scalar diquark correlation and hence $d_v/u_v|_{x=1}\neq 0$.  Indeed, the antithesis of scalar-diquark-only models is dominance of axial-vector diquark correlations, in which case Eq.\,\eqref{dvuvF1result} produces $d_v/u_v|_{x=1}=2$.
A dynamical equivalence between the scalar and axial-vector diquark correlations provides another special instance.  In this case $P_1^{p,s}=P_1^{p,a}$
and hence $d_v/u_v|_{x=1}=1/2$. 

\begin{table}[t]
\begin{center}
\begin{tabular}{lrrrrrrr}\hline
    & $\frac{F_2^n}{F_2^p}$ & $\frac{d}{u}$ & $\frac{\Delta d}{\Delta u}$
    & $\frac{\Delta u}{u}$ & $\frac{\Delta d}{d}$ & $A_1^n$ & $A_1^p$\\\hline
%
DSE-1
& $0.49$ & $0.28$ & $-0.11$ & 0.65 & $-0.26$ & 0.17 & 0.59 \\
DSE-2& $0.41$ & $0.18$ & $-0.07$ & 0.88 & $-0.33$ & 0.34 & 0.88\\\hline
$0_{[ud]}^+$ & $\frac{1}{4}$ & 0 & 0 & 1 & 0 & 1 & 1 \\[0.5ex]
NJL & $0.43$ & $0.20$ & $-0.06$ & 0.80 & $-0.25$ & 0.35& 0.77 \\[0.5ex]
SU$(6)$ & $\frac{2}{3}$ & $\frac{1}{2}$ & $-\frac{1}{4}$ & $\frac{2}{3}$ & $-\frac{1}{3}$ & 0 & $\frac{5}{9}$ \\[0.5ex]
CQM & $\frac{1}{4}$ & 0 & 0 & 1 & $-\frac{1}{3}$ & 1 & 1 \\[0.5ex]
pQCD & $\frac{3}{7}$ & $\frac{1}{5}$ & $\frac{1}{5}$ & 1 & 1 & 1 & 1 \\\hline
\end{tabular}
\caption{\label{tab:a}
Selected predictions for the $x=1$ value of the indicated quantities.
The DSE results are computed as described herein: DSE-1 (also denoted ``DSE realistic'' below) indicates use of the momentum-dependent dressed-quark mass-function in Ref.\,\protect\cite{Cloet:2008re}; and DSE-2 (also denoted ``DSE contact'') corresponds to predictions obtained with a contact interaction \protect\cite{Roberts:2011wy}.
The next four rows are, respectively, results drawn from Refs.\,\protect\cite{Close:1988br,Cloet:2005pp,Hughes:1999wr,Isgur:1998yb}.
The last row, labeled ``pQCD,'' expresses predictions made in Refs.\,\protect\cite{Farrar:1975yb,Brodsky:1994kg}, which are actually model-dependent: they assume an SU$(6)$ spin-flavour wave function for the proton's valence-quarks and the corollary that a hard photon may interact only with a quark that possesses the same helicity as the target.
}
\end{center}
\end{table}

Turning toward realistic scenarios, two distinct Faddeev equation kernels were considered in Ref.\,\cite{Wilson:2011aa}; viz., that connected with the nucleon form factor predictions in Ref.\,\cite{Cloet:2008re}, which corresponds to the realistic case of a momentum-dependent dressed-quark mass function, and that based upon the DSE treatment of a vector$\,\otimes$vector contact interaction \cite{Roberts:2011wy}, which produces a momentum-independent dressed-quark mass.  Results from these dynamical calculations, computed using the probability values specified below, are listed in Table~\ref{tab:a}.

We will now deduce formulae analogous to Eq.\,\eqref{dvuvF1result} for the spin-dependent valence-quark distributions at $x=1$.  To that end, consider the general Faddeev amplitude in Ref.\,\cite{Cloet:2008re}, which expresses the relative-momentum dependence of quark-diquark configurations within the nucleon, where the diquarks are either scalar or axial-vector correlations: $\Psi = \Psi_{0^+} + \Psi_{1^+}$.

Recall that $P_1^{p,s}$ is the probability for finding a $u$-quark bystander in association with a scalar $ud$-diquark correlation in the proton.  Owing to Poincar\'e covariance, this term expresses a sum of quark-diquark angular momentum $L^{u[ud]}=0$ and $L^{u[ud]}=1$ correlations within the nucleon.  With $L^{u[ud]}=0$, the bystander quark carries all the nucleon's spin.  On the other hand, the $L^{q[ud]}=1$ correlation contributes to both the parallel and antiparallel alignment probabilities of the bystander quark: $2 [ud]_{L_z^{u[ud]}=1} u_{\downarrow} \oplus [ud]_{L_z^{u[ud]}=0} u_{\uparrow}$.  The relative strength of these terms is fixed by solving the Faddeev equation and expressed thereafter in the Faddeev amplitude: $\Psi_{0^+} \sim \psi_{L=0} + \psi_{L=1}$, so that, converting the amplitude to probabilities,
\begin{equation}
\label{Pscalar}
\begin{array}{ll}
P_1^{p,s} = P^{p,s}_{u_\uparrow} + P^{p,s}_{u_\downarrow}, & \\
P^{p,s}_{u_\uparrow} = \psi_{L=0}^2 + 2 \psi_{L=0} \psi_{L=1}+ \frac{1}{3} \psi_{L=1}^2, &
P^{p,s}_{u_\downarrow} =  \frac{2}{3} \psi_{L=1}^2.\;
\end{array}
\end{equation}
Following Ref.\,\cite{Wilson:2011aa}, one finds $\psi_{L=0}=0.88$, $\psi_{L=1}=0$ because that treatment of the contact interaction produces a momentum-independent nucleon Faddeev amplitude.  On the other hand, the Faddeev equation used in Ref.\,\cite{Cloet:2008re}, based upon a momentum-dependent dressed-quark mass function, yields $\psi_{L=0}=0.55$, $\psi_{L=1}=0.22$.  (These values were computed by combining results in Refs.\,\cite{Cloet:2008re,Cloet:2007piS}.)

The probability for finding a quark bystander in association with an axial-vector diquark correlation in the proton is $P_1^{p,a}$.  In this case the bystander quark can be either a $u$- or $d$-quark.  Confronted with the fact that the presence of axial-vector diquarks entails that the proton's Faddeev amplitude expresses $S$-, $P$- and $D$-wave quark-diquark orbital angular momentum correlations, one might become confused when attempting to determine the association between these components and the parallel vs.\ antiparallel alignment probabilities of the bystander quark.  The answer is simple, however.  It does not matter which $L^{q\{qq\}}$-wave one considers, one must only couple $L^{q\{qq\}}$ and $J^{\{qq\}}$ to a total angular momentum, ${\cal J}^{\{qq\}}$, which, when combined with $J^q=1/2$ for the bystander quark, yields a spin-up proton.  This means that $L^{q\{qq\}}+ J^{\{qq\}}$ must be either $|{\cal J},{\cal J}_z\rangle = |1,1\rangle$ or $|1,0\rangle$.  Accounting now for both the flavour and angular momentum Clebsch-Gordon coefficients, one arrives at
\begin{equation}
\label{Paxial}
\begin{array}{ll}
P^{p,a}_{u_\uparrow} = \frac{1}{9} P_1^{p,a},&
P^{p,a}_{u_\downarrow} = \frac{2}{9} P_1^{p,a}, \\
P^{p,a}_{d_\uparrow} = \frac{2}{9} P_1^{p,a}, &
P^{p,a}_{d_\downarrow} = \frac{4}{9} P_1^{p,a}.
\end{array}
\end{equation}
The two DSE interactions under consideration herein yield, respectively,  $P_1^{p,a}=0.25$, $0.22$ \cite{Cloet:2008re,Wilson:2011aa}.

The remaining possibility is the mixed configuration, associated with $P_1^{p,m}$, which describes the probability of scalar--axial-vector diquark mixing.  As noted above, the hard piece of this configuration contributes twice as much to $u_v$ as $d_v$.  Moreover, the quark that leaves the scalar $[ud]$-diquark correlation, to join with the bystander and form the axial-vector diquark, has equal parallel and antiparallel alignment probabilities.  It follows that
\begin{equation}
\label{Pmixed}
\begin{array}{ll}
P^{p,m}_{u_\uparrow} = \frac{1}{3} P_1^{p,m},&
P^{p,m}_{u_\downarrow} = \frac{1}{3} P_1^{p,m},\\
P^{p,m}_{d_\uparrow} = \frac{1}{6} P_1^{p,m},&
P^{p,m}_{d_\downarrow} = \frac{1}{6} P_1^{p,m}.
\end{array}
\end{equation}
The DSE interactions under consideration herein yield, respectively,  $P_1^{p,m}=0.14$, $0$ \cite{Cloet:2008re,Wilson:2011aa}.

Combining Eqs.\,\eqref{Pscalar}--\eqref{Pmixed}, we find:
\begin{equation}
\label{proball}
\begin{array}{l}
P^p_{u_\uparrow} = \psi_{L=0}^2 + 2 \psi_{L=0} \psi_{L=1}+ \frac{1}{3} \psi_{L=1}^2 + \frac{1}{9} P_1^{p,a} + \frac{1}{3} P_1^{p,m},\\
P^p_{u_\downarrow} = \frac{2}{3} \psi_{L=1}^2 + \frac{2}{9} P_1^{p,a} + \frac{1}{3} P_1^{p,m},\\
P^p_{d_\uparrow} = \frac{2}{9} P_1^{p,a} + \frac{1}{6} P_1^{p,m}, \\
P^p_{d_\downarrow} = \frac{4}{9} P_1^{p,a} + \frac{1}{6} P_1^{p,m}.
\end{array}
\end{equation}
With $\Delta q := (P^p_{q_\uparrow} - P^p_{q_\downarrow})/(P^p_{q_\uparrow} + P^p_{q_\downarrow})$ and standard expressions for $A_1^{p,n}$ (e.g., Ref.\,\cite{Hughes:1999wr}), Eqs.\,\eqref{proball} produce the results in Table~\ref{tab:a}.  Plainly, $(-1/3)<(\Delta d/d)< (-2/9)$ for $0\leq P_1^{p,m}\leq P_1^{p,a}$.
\smallskip

\hspace*{-\parindent}\textbf{4 Experimental Status: \mbox{\boldmath $F_2^n/F_2^p$} and \mbox{\boldmath $d/u$}}.
%
%
The ratio of the neutron structure function, $F_2^n$, to the proton structure function, $F_2^p$, is particularly interesting.  Within the parton model, on the far valence domain:
\begin{equation}
\label{F2nF2pratio}
\frac{F_2^n}{F_2^p} \stackrel{x \simeq 1}{=} \frac{1 + 4(d_v/u_v)}{4 + (d_v/u_v)}.
\end{equation}
Thus a measurement of the neutron and proton structure functions at large-$x$ provides a determination of the $d_v/u_v$ ratio.  However, while proton and deuteron DIS data are well measured at reasonably high $x$, the extraction of the neutron structure function at very high $x$ from DIS data on the deuteron is problematic.  The central difficulty is that the extraction of $F_2^n/F_2^p$ at high $x$ is sensitive to the poorly known high-momentum components of the deuteron wave function \cite{Arrington:2011xs}.

\begin{figure}[t]
\includegraphics[clip,angle=-90,width=0.48\textwidth]{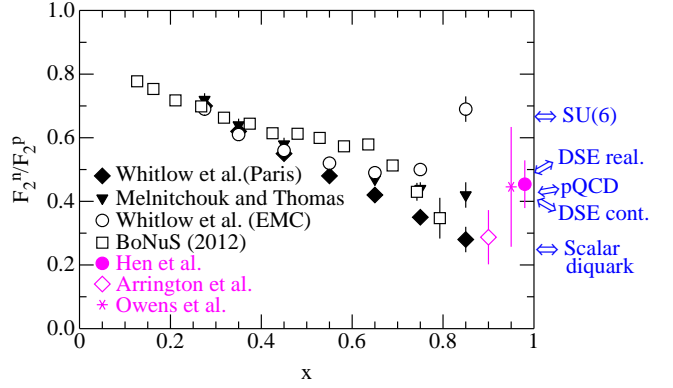}
\caption{\label{f2nratio} $F_2^n/F_2^p$ as a function of $x$.  Results from five extraction methods are shown \protect\cite{Whitlow:1991uw,Melnitchouk:1995fc,Arrington:2008zh,Hen:2011rt,Arrington:2011qt,Owens:2012bv} along with selected predictions from Table~\protect\ref{tab:a}. N.B.\ In all figures, DSE~realistic corresponds to DSE-1 and DSE~contact to DSE-2.
}
\end{figure}

To see this, we note that many extractions of the neutron-proton structure function ratios have been performed \cite{Whitlow:1991uw,Melnitchouk:1995fc,Arrington:2008zh,Hen:2011rt,Arrington:2011qt,Owens:2012bv}.  They are summarised in Fig.\,\ref{f2nratio}, with the three most recent inferences indicated by the points with error bars near $x=1$: there is a large uncertainty in the ratio for $x \gtrsim 0.6$.  (See also Fig.\,25 in Ref.\,\cite{Holt:2010vj}.)  New experimental methods are necessary in order to place tighter constraints on the far valence domain.  A primary goal should be to empirically eliminate two of the three materially different theoretical predictions; viz., to unambiguously distinguish between $F_2^n/F_2^p=1/4$, $F_2^n/F_2^p\approx 1/2$, $F_2^n/F_2^p =2/3$.  (N.B.\ The value $F_2^n/F_2^p=1/4$ is already being challenged empirically \cite{Weinstein:2010rt}.)

In this connection, two new experiments \cite{Bueltmann:2006,Petratos:2006} will focus on providing data up to $x \approx 0.85$.  Since much of the uncertainty can be traced to the poorly known short-range part of the deuteron wave function, $\psi_D$, the JLab BoNuS Collaboration has performed \cite{Baillie:2011za} an experiment where a very low energy spectator proton from the deuteron can be detected in coincidence with a DIS event from the neutron in the deuteron.  In this way, one can restrict the data to a region where the well-known long-range part of $\psi_D$ dominates the process.  An interesting variant of this approach is to use an EIC with, e.g., an 8$\,$GeV electron beam impinging on a deuteron beam of 30$\,$GeV in energy.  The forward going $\sim 30\,$GeV proton would be detected at very small angles in coincidence with a DIS event from the neutron.  Simulations suggest that this should be feasible \cite{Accardi:2010}.

Another method is to perform deep inelastic scattering from the mirror nuclei $^3$He and $^3$H over a broad range in $x$ \cite{Afnan:2000uh,Bissey:2000ed,Afnan:2003vh,Petratos:2006}.  Theoretical calculations indicate that nuclear effects cancel to a high degree in extracting the $F_2^n/F_2^p$ ratio from these two nuclei.  This experiment would also be useful in determining the EMC effect in the mass-three system \cite{Hen:2013oha}.

Finally, parity violating DIS can avoid the problem encountered with neutrons bound in nuclei; and, moreover, parity-violating DIS from the proton is sensitive to the $d/u$ ratio on $x \lesssim 0.7$ \cite{Souder:2010}.

\smallskip

%
\hspace*{-\parindent}\textbf{5 Experimental Status: longitudinally polarised DIS}.
It is evident from Table~\ref{tab:a} that measurements of the longitudinal asymmetries in DIS provide a sensitive additional constraint on models of nucleon structure.  Numerous experiments and extractions aimed at determining nucleon longitudinal spin structure functions have been performed \cite{Kuhn:2008sy,Alekseev:2010ub} 

\begin{figure}[t]
\includegraphics[clip,angle=-90,width=0.4\textwidth]{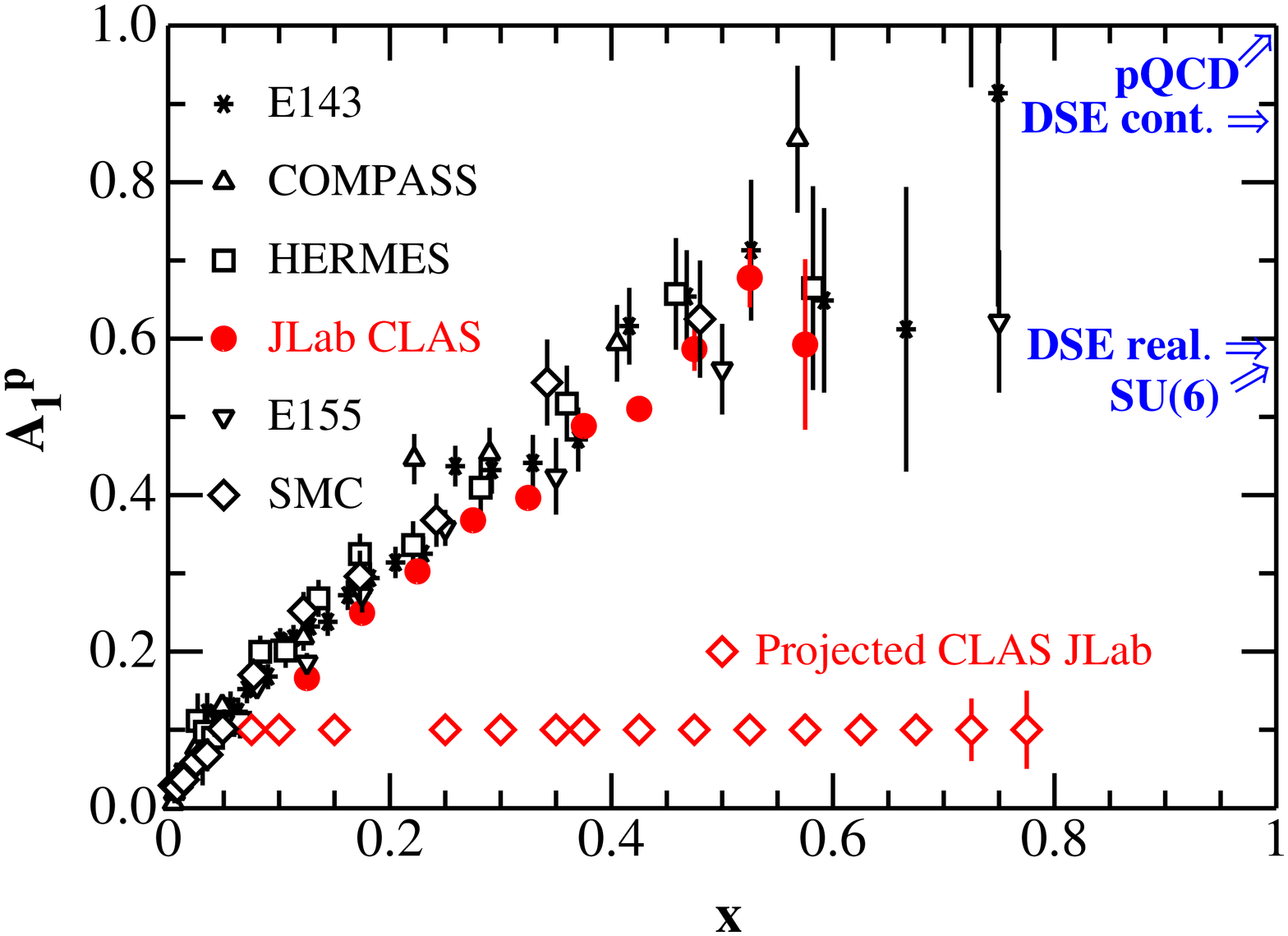}
\includegraphics[clip,angle=-90,width=0.4\textwidth]{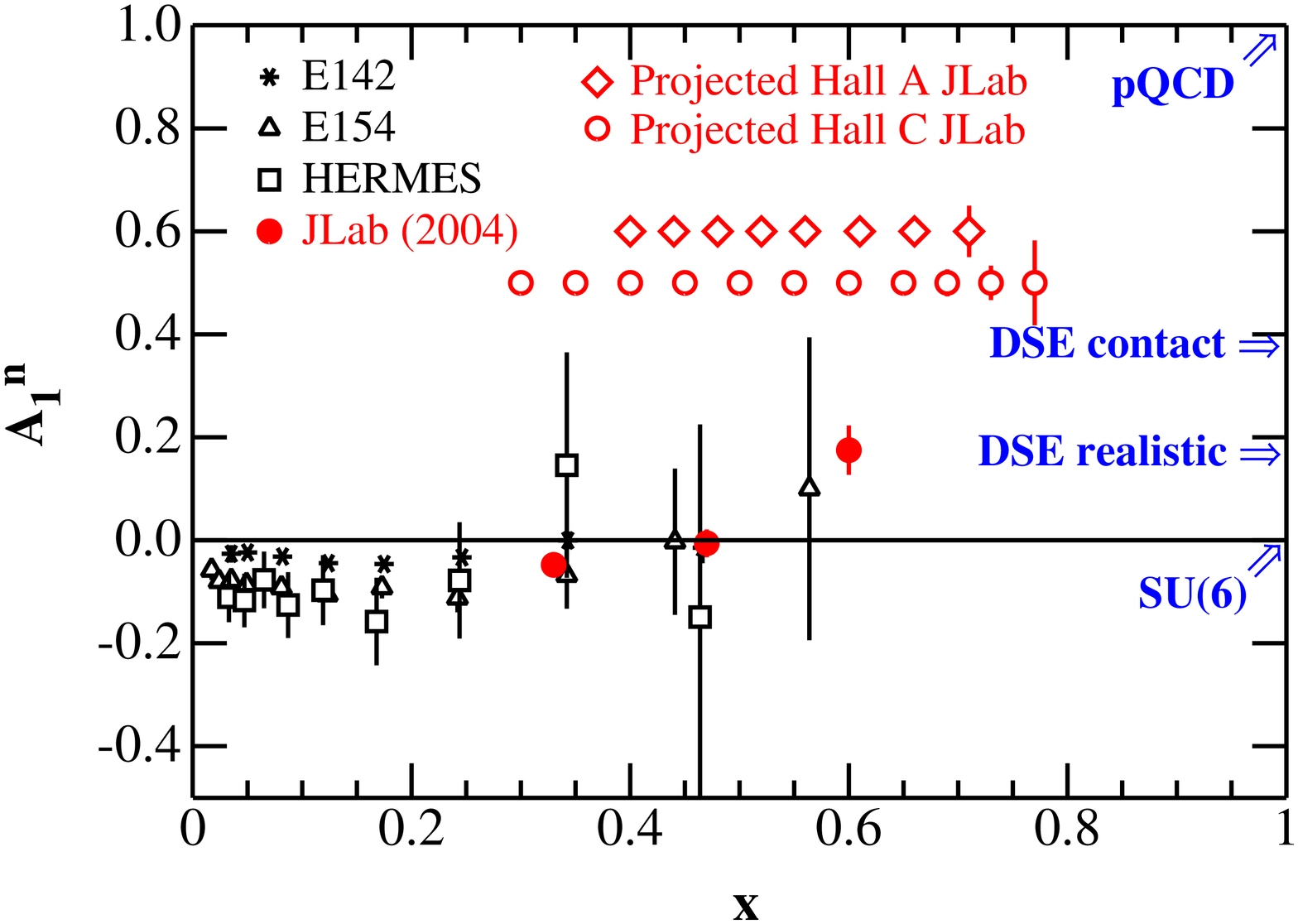}
\caption{\label{a1p}
\emph{Upper panel}.  Existing and projected measurements of the proton's longitudinal spin asymmetry as a function of $x$ (statistical errors only), along with selected predictions from Table~\protect\ref{tab:a}.  \emph{Lower panel}.  Same for the neutron.  N.B.\ We only display $A_1^n$ data obtained from polarised $^3$He targets.}
\end{figure}

Existing measurements of $A_1^p$ are summarised in the upper panel of Fig.\,\ref{a1p}.  They fail to discriminate between the model predictions in Table~\ref{tab:a}.  As indicated in Fig.\,\ref{a1p}, however, a new experiment \cite{kuhn:2006} will extend the results up to $x \approx 0.8$ with a projected error that promises to add a capacity for discrimination.


The status of existing data for $A_1^n$ is shown in the lower panel of Fig.\,\ref{a1p}.  The data extend only to $x \approx 0.6$ and, as evident from a comparison with Table~\ref{tab:a}, place little constraint on descriptions of the nucleon.  New experiments proposed at JLab \cite{Zheng:2006,Liyanage:2006} are expected to provide results up to $x \approx 0.75$, as indicated in the lower panel of Fig.\,\ref{a1p}.  They promise to enable discrimination between the pQCD model and other predictions.
\smallskip

\hspace*{-\parindent}\textbf{6 Epilogue}.
A key element in the international effort to understand how the interactions between dressed-quarks and -gluons create hadron bound-states, and how these interactions emerge from QCD, is the program to chart and explain the behavior of parton distribution functions on the far valence domain.  Of particular importance are the ratios of distribution functions on $x\simeq 1$.  Such ratios are an unambiguous, scale invariant, nonperturbative feature of QCD and are therefore a keen discriminator between frameworks that claim to explain hadron structure.  In this connection, our analysis has stressed that empirical results for nucleon longitudinal spin asymmetries on $x\simeq 1$ promise to add greatly to our capacity for discriminating between contemporary pictures of nucleon structure.

\smallskip

\hspace*{-\parindent}\textbf{Acknowledgments}.
We are grateful for insightful comments from I.\,C.~Clo\"et.
CDR acknowledges support from an Helmholtz Association \emph{International Fellow Award}.
Work otherwise funded by:
Department of Energy, Office of Nuclear Physics, contract no.~DE-AC02-06CH11357;
and For\-schungs\-zentrum J\"ulich GmbH.


\end{document}